\documentclass[twocolumn]{aastex}

\usepackage[T1]{fontenc}
\usepackage{ae,aecompl}

\usepackage{graphicx}
\usepackage{epstopdf}
\usepackage{amsmath}

\usepackage{natbib}

\begin{document}

\title{Modeling the afterglow of the possible Fermi-GBM event associated with GW150914}


\author{Brian J. Morsony\altaffilmark{1}}
\affil{Dept. of Astronomy, University of Maryland, 1113 Physical Sciences Complex, College Park, MD, 20742-2421, USA}

\author{Jared C. Workman}
\affil{Dept. of Physical and Environmental Sciences, Colorado Mesa University, Grand Junction, CO, 81501, USA}

\author{Dominic M. Ryan}
\affil{Dept. of Astronomy, University of California, Berkeley, 501 Campbell Hall \#3411, Berkeley, CA, 94720-3411, USA}

\altaffiltext{1}{morsony@astro.umd.edu}


\begin{abstract}

We model the possible afterglow of the {\it Fermi} GBM event associated with LIGO detection GW150914, under the assumption that the gamma-ray are produced by a short GRB-like relativistic outflow.  
We model GW150914-GBM as both a weak, on-axis short GRB and normal short GRB seen far off axis.
Given the large uncertainty in the position of GW150914, we determine that the best chance of finding the afterglow is with ASKAP or possibly the MWA, with the flux from an off-axis short GRB reaching 0.2 - 4 mJy (0.12 - 16 mJy) at 150 MHz (863.5 MHz) by 1 - 12 months after the initial event.  
At low frequencies, the source would evolve from a hard to soft spectrum over several months.  
The radio afterglow would be detectable for several months to years after it peaks, meaning the afterglow may still be detectable and increasing in brightness {\bf NOW} (mid-July 2016).
With a localization from the MWA or ASKAP, the afterglow would be detectable at higher radio frequencies with the ATCA and in X-rays with {\it Chandra} or XMM.

\end{abstract}


\section{Introduction}

GW150914 is the first gravitational wave source detected by LIGO \citep{abbott16a}, and is the merger of two approximately $30$~M$_{\sun}$ black holes.
The merger occurred at a distance of $410^{+160}_{-180}$~Mpc ($z=0.09^{+0.03}_{-0.04}$) in the southern hemisphere \citep{abbott16a}.
Although the merger of two large black holes was not expected to produce a bright electromagnetic signature \citep{connaughton16}, the {\it Fermi} Gamma-ray Burst Monitor (GBM) found a 1 second increase in gamma-ray emission $0.4$ seconds after the LIGO detection, and located in the same region of the sky \citep{connaughton16}.
INTEGRAL did not detect an event associated with GW150914 \citep{savchenko16}, in tension with the GBM results.

Assuming the GBM detection is associated with GW150914, it would have an isotropic luminosity of $E_{iso}=1.8^{+1.5}_{-1.0}\times10^{49}$~erg, an order of magnitude fainter than any other short-GRB analyzed \citep{connaughton16,wanderman15}.
Typical short GRB energies range from $E_{iso}=10^{50}-10^{52}$ \citep{fong15}.
The properties of the {\it Fermi} GBM detection are broadly consistent with a short GRB, although it would be significantly harder than the typical $E_{peak}$ - $E_{iso}$ relation for short GRBs \citep{li16}.
It is possible that GW150914 produced either an under-luminous short-GRB, a poorly collimated jet, or that it is a typical short GRB but seen far off-axis, so that the prompt emission directed towards Earth was faint.

GW150914 occurred in the southern hemisphere, but was poorly localized, with the combined LIGO and {\it Fermi} GBM error boxes covering $199$~deg$^2$ \citep{connaughton16}.  
This made it very difficult to localized the early, rapidly fading X-ray or optical afterglow. 
Non-detections have been reported by {\it Swift} XRT and UVOT \citep{evans16}, the Dark Energy Camera \citep{soares-santos16}, and Pan-STARRS and PESSTO \citep{smartt16}.


In the radio, a GRB's afterglow produces a persistent source, but is initially self-absorbed, taking weeks to months to reach peak luminosity.
The best prospect for finding the afterglow would be with a wide-field instrument, such as the Murchinson Widefield Array (MWA) , a low-frequency ($80$ - $300$ MHz) radio array \citep{tingay13}, or the ASKAP Boolardy Engineering Test Array (BETA) at $713.5$ - $1013.5$~MHz \citep{hobbs16}, both located in Australia.
MWA is sensitive to sources brighter than about $\sim6$~mJy before becoming confusion limited at $150$~MHz \citep{tingay13}, and ASKAP is currently sensitive down to $\sim3$~mJy at $862.5$~MHz \citep{gcn18363}.
The Australia Compact Telescope Array (ATCA), at $1.1$ - $3.1$ GHz, is sensitive to much fainter sources ($0.04$mJy in 10 minute integration), but has a much smaller field of view ($0.4$ to $0.05$~deg$^2$ vs. $600$~deg$^2$ for MWA and $30$~deg$^2$ for ASKAP), making covering the large possible area for GW150914 difficult.  It would, however, be able to follow up any candidate afterglow sources.


The best way to determine if LIGO event GW150914 and the {\it Fermi} GBM detection are associated would be to find an afterglow that can be localized to a host galaxy.

We model the possible afterglow of GW150914-GBM as a relativistic blastwave expanding into the surrounding medium, and emitting synchrotron radiation.  
For a review of afterglow modeling, see \citet{piran04}.

The hydrodynamics of a GRB afterglow are typically modeled either using a semi-analytic solution for an expanding spherical blastwave, or with direct hydrodynamical fluid simulations.
Since a GRB is typically only detected if the Earth is within the jet opening angle, and the outflow is initially highly relativistic, a natural model is a spherical, ultra-relativistic Blandford-McKee blastwave \citep{blandford76}, as in, e.g., \citet{granot99,vanEerten09}.  
More accurate semi-analytic models of spherical and on-axis jet afterglows have also been used, such an an interpolation between a Blandford-McKee and Sedov-Taylow solution in \citet{decolle12}, and an analytic fit to 1D hydrodynamic simulations in \citet{leventis12}.


Semi-analytic models for an off-axis observer are also possible \citep[e.g.][]{granot02,waxman04,rossi04,morsony07,vanEerten10}, but these models all assume an ultra-relativistic Blandford-McKee blastwave.  
Hydrodynamical simulations of on- and off-axis GRB afterglows have been carried out \citep[e.g.][]{granot02,cannizzo04,vanEerten10,vanEerten11,decolle12,decolle12b}.  
Although very accurate, full hydrodynamic simulations are computationally expensive and can only be used to directly model a small number of cases.

Our modeling approach, described in section~\ref{sect:code}, is to use a semi-analytic off-axis blastwave model that smoothly transitions between an ultra-relativistic and non-relativistic expansion, following \citet{decolle12}.  
%
%
In section~\ref{sect:results} we use our afterglow code to make specific predictions for possible afterglow light curves of GW150914, given a range of possible model parameters.
In section~\ref{sect:conclusions} we discuss under what conditions an afterglow of GW150914 might be detectable.



\section{Methods}
\label{sect:code}

We carry out modeling of the afterglow using the Trans-Relativistic Afterglow Code (TRAC).
TRAC is able to solve a semi-analytic model of the emission of a relativistic fireball with an arbitrary angular distribution of energy and at an arbitrary observer angle (i.e. relative to a jet axis).
TRAC is based on the methodology of \citet{granot99} and began development for use in \citet{morsony09}.
A full description of TRAC will be published in Morsony et al. (in prep.).

Beginning with an energy distribution and observer angle, at a given observer time, TRAC first calculates the size of the blastwave seen by the observer, along with the values of density, pressure, and Lorentz factor inside this shell.  TRAC treats the evolution along each radial direction from the center of the blast as an independently evolving portion of a spherical blastwave (no transverse mixing).  
Based on the interpolation of \citet{decolle12} (eqn.~\ref{eqn:decolle}), we smoothly transitions between the ultra-relativistic Blandford-McKee phase \citep{blandford76} and the non-relativistic Sedov-Taylor phase.  
This is important for radio observations, which peak during the mildly relativistic transition phase.
At any time, the total energy of a spherical blastwave satisfies
\begin{equation}
E=R^{3-k}\beta^{2}\Gamma^{2}\rho_{0} c^{2}\left[\frac{8\pi}{17-4k}\beta^{2}+\frac{(5-k)^{2}}{4\alpha_{k}}(1-\beta^{2})\right]
\label{eqn:decolle}
\end{equation}
where $E$ is the blast energy, $R$ is the shock radius, $\Gamma$ is the Lorentz factor of the shock, $\beta$ is the shock velocity over $c$, and, for a constant density external medium, $\rho_{0}$ is the external mass density, $k=0$, and $\alpha_{k}=6.5159$.

Using the 3-dimensional grid of fluid values, TRAC then computes the integrated emission from the blastwave as a function of frequency.  For the current version of TRAC, we assume the emission is due to synchrotron radiation parameterized by $\epsilon_e$, the fraction of energy in electrons, $\epsilon_B$, the fraction of energy in the magnetic field, and $p$, the spectral index of the electron energy distribution.  
TRAC solves the radiative transfer equation including local synchrotron cooling and synchrotron self-absorption.  
All models considered are in the slow cooling regime.

For the models presented here, we assume the shocked gas has an adiabatic index of $\gamma_{ad}=4/3$ and that the structure behind the shock front follows that of a relativistic shock, i.e.
\begin{equation}
\rho(\chi)=\rho_{f}\chi^{-(7-2k)/(4-k)}
\end{equation}
\begin{equation}
P(\chi)=P_{f}\chi^{-(17-4k)/(12-3k)}
\end{equation}
\begin{equation}
\gamma(\chi)=\sqrt{(\gamma_f-1)^2/\chi}+1
\end{equation}
where $\chi=1+4(m+1)(1-r/R)\gamma_f^2$ is a self-similarity variable, $\rho_{f}$, $P_{f}$, and $\gamma_{f}$ are density, pressure, and Lorentz factor of fluid immediately behind the shock front, and $r$ is radius at any position behind the shock.  For an impulsive explosion with $k=0$, $m=3$.
We also assume that the jet maintains a constant opening angle and does not spread out.
Both of these approximations will break down in the non-relativistic limit.  
In this limit, using a Sedov-Taylor rather than Blandford-McKee internal structure would increase the flux density by up to a factor of 4.5.  Allowing the jet to spread reduces the flux density by a factor of 1.5 to 2.  The net effect is that our models under-predicts the late-time flux by up to a factor of 2.5.

For all of our afterglow models, we assume the event is located at $D=410$~Mpc, with $\epsilon_e=0.1$, $\epsilon_B=0.01$, $p=2.5$ the observer-directed kinetic energy is $E_{iso}=1.8\times10^{49}$~erg, and the blastwave is expanding into a constant density environment (ISM).
We use typical values for the density around observed short GRBs of $n_H=10^{-2}$, $10^{-1}$, and $10^{0}$~cm$^{-3}$ \citep{fong15}.

For these values, the self-absorption frequency is initially 
\begin{equation}
\nu_{a}=2.39\times10^{8}E_{51}^{1/5}n_{-1}^{3/5}\rm{~Hz}
\end{equation}
\citep{panaitescu00} until the emission peak frequency equals the self-absorption frequency, after which $\nu_{a} {\propto} t^{-19/26}$.

\section{Results}
\label{sect:results}

We model the possible afterglow of GW150914 with two different basic assumptions: 1) that GW150914-GBM is an under-luminous short GRB that is isotropic or seen on-axis, or 2) that it is a typical short-GRB seen far off-axis, outside the prompt emission cone of the main jet, but with an Earth-directed component that accounts for the {\it Fermi} GBM detection.
For each of these assumption, we model the afterglow with different ISM densities, and, for the off-axis models, different observer angles and jet energies.
A full list of model parameters is given in table~\ref{table:models}.

\begin{table*}
\centering
\caption{Model Data \label{table:models}}


\begin{tabular}{l||r|l|l|l|l|l|l}
 \hline
   Model & $E_{iso}$ & $n_H$ & $\theta_{obs}$ & Peak flux\tablenotemark{a} & Peak time\tablenotemark{a} & Peak flux\tablenotemark{b} & Peak time\tablenotemark{a} \\ 
   ~ & (erg) & (cm$^{-3}$) & (deg) & (mJy) & (weeks) & (mJy) & (weeks) \\ \hline
  iso            &  1.8e49     &  0.1  & 0            &  $0.11$  &  $3$    & $0.10$  & $1$ \\
  iso hi         &  1.8e49     &  1    & 0            &  $0.23$  &  $7$    & $0.34$  & $1$ \\  
  iso lo         &  1.8e49     &  0.01 & 0            &  $0.03$  &  $3$    & $0.03$  & $1$ \\
  on-axis        &  1.8e49     &  0.1  & 0            &  $0.03$  &  $0.6$  & $0.04$  & $0.3$ \\
  on-axis hi     &  1.8e49     &  1    & 0            &  $0.03$  &  $0.2$  & $0.10$  & $0.2$ \\
  on-axis lo     &  1.8e49     &  0.01 & 0            &  $0.01$  &  $0.7$  & $0.02$  & $0.4$ \\
  51.-1.30       &  1.e51      &  0.1  & 30           &  $0.34$  &  $16$   & $0.27$  & $9$ \\
  51.-1.20       &  1.e51      &  0.1  & 20           &  $0.60$  &  $9$    & $0.66$  & $4$ \\
  51.-1.40       &  1.e51      &  0.1  & 40           &  $0.20$  &  $22$   & $0.12$  & $14$ \\
  51.0.30        &  1.e51      &  1    & 30           &  $0.46$  &  $17$   & $1.1$   & $5$ \\
  51.-2.30       &  1.e51      &  0.01 & 30           &  $0.10$  &  $22$   & $0.06$  & $14$ \\
  51.0.20        &  1.e51      &  1    & 20           &  $0.55$  &  $17$   & $2.1$  & $3$ \\
  52.-1.30       &  1.e52      &  0.1  & 30           &  $2.3$   &  $38$   & $2.4$  & $19$ \\
  52.0.30        &  1.e52      &  1    & 30           &  $2.1$   &  $56$   & $8.2$  & $13$ \\
  52.-1.20       &  1.e52      &  0.1  & 20           &  $4.0$   &  $24$   & $6.0$  & $10$ \\
  52.0.20        &  1.e52      &  1    & 20           &  $2.5$   &  $52$   & $16.$  & $8$ \\
\hline 
      \end{tabular}

\tablenotetext{a}{at $150$~MHz}
\tablenotetext{b}{at $863.5$~MHz}

   

\end{table*}

\subsection{On-axis, under-luminous short GRB}

For our first set of models, we assume GW150914-GBM is an under-luminous short GRB.
We assume the GRB is either a beamed jet seen on-axis ($\theta_{obs}=0$) with a half-opening angle of $\theta_j=10$~degrees, or is an isotropic explosion ($\theta_j=90$~degrees).
We model each case with 3 different background densities of $n_H=10^{-2}$, $10^{-1}$, and $10^{0}$~cm$^{-3}$.

\begin{figure}
\centering
\includegraphics[width=0.7\columnwidth]{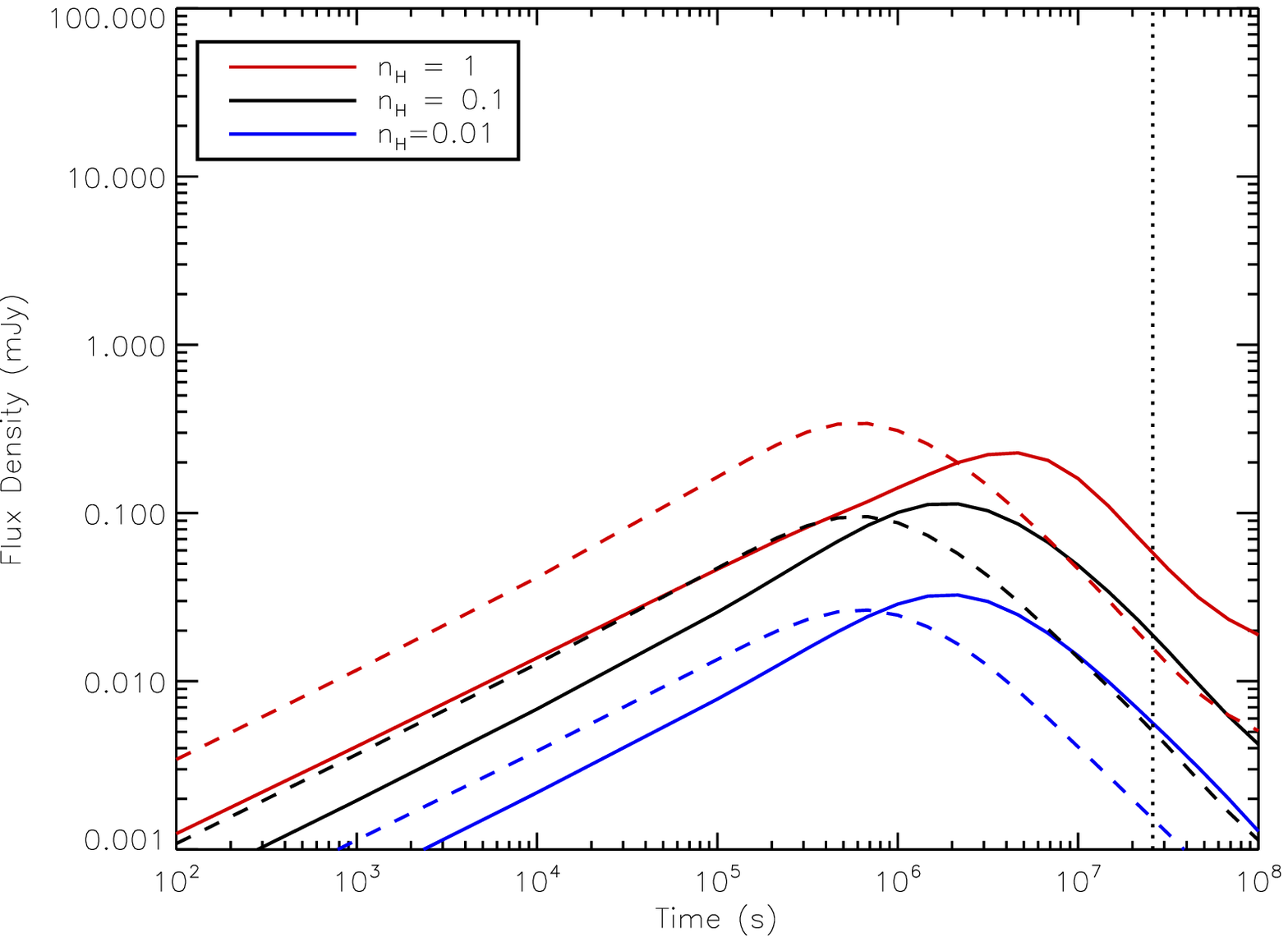}
\includegraphics[width=0.7\columnwidth]{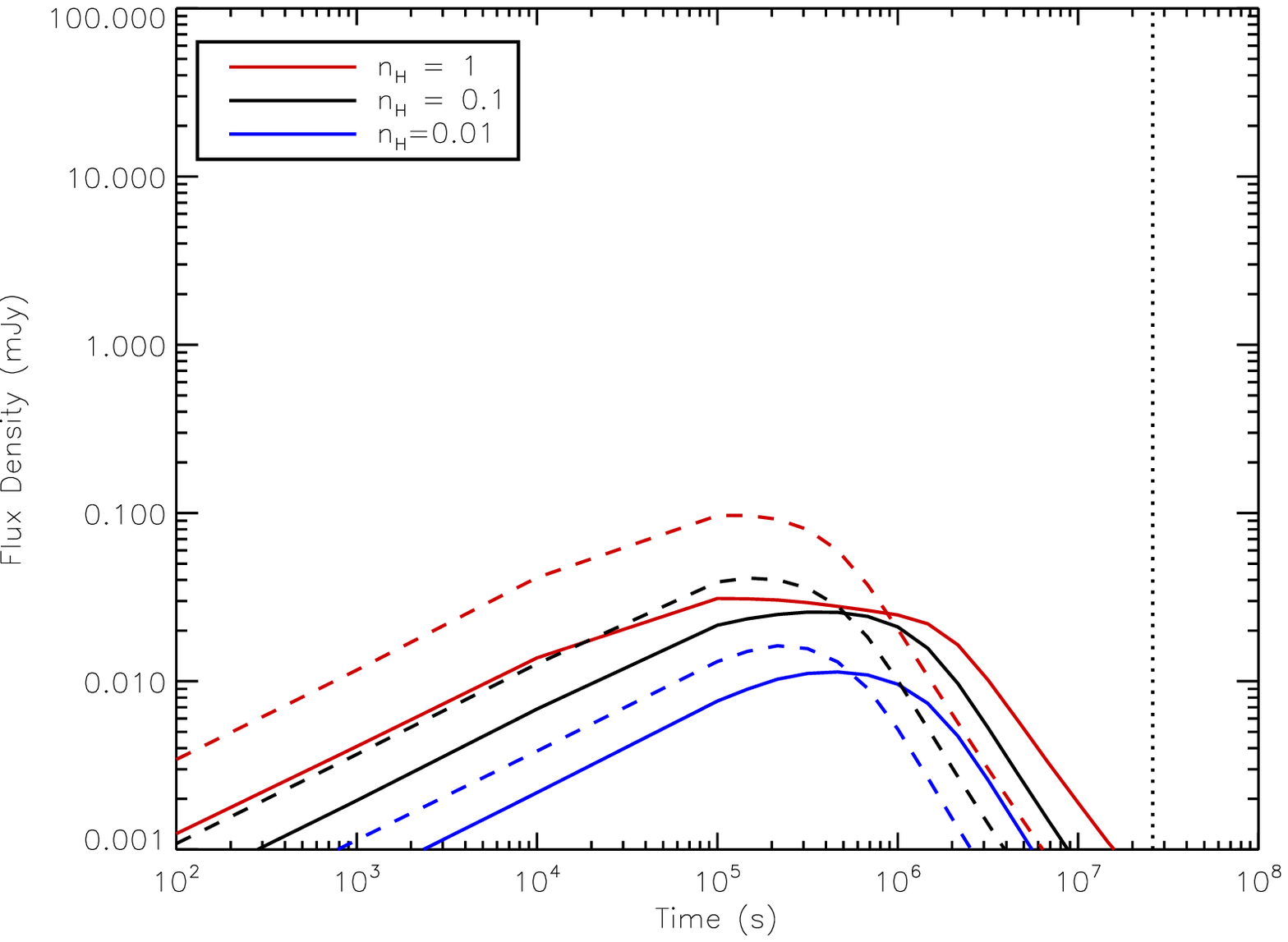}
\caption{
{\it Top}: Modeled afterglow flux assuming an isotropic energy distribution of $1.8\times10^{49}$~erg at $150$~MHz (solid lines) and $863.5$~MHz (dashed lines).  Models are for external densities of $n_H=1$~cm$^{-3}$ (red), $0.1$~cm$^{-3}$ (black) and $0.01$~cm$^{-3}$ (blue).  
Dotted vertical line is set 10 months after the initial event, corresponding to July 10, 2016 for GW150914, the approximate publication date of this paper.
{\it Bottom}: Same as top panel, but assuming the energy is in jet with $\theta_j=10$~degrees, with $E_{iso}=1.8\times10^{49}$~erg and directed towards Earth.
}
\label{fig:sphere_jet}
\end{figure}

Fig.~\ref{fig:sphere_jet} (top panel) plots the flux density predicted at Earth at $150$~MHz (solid lines) and $863.5$~MHz (dashed lines) for the isotropic cases.  
For GW150914, the vertical dotted line corresponds July 10, 2016, the approximate publication date of this paper.
Higher external density produces a brighter radio flux. 
It does not change the time of the peak flux, except for the highest density model at $150$ MHz, where the blastwave is still optically thick.
In this case the peak is delayed from 3 weeks after the GRB to 7 weeks.

In the jetted cases (fig.~\ref{fig:sphere_jet}, bottom panel), higher densities lead to an earlier time of peak emission, as the blastwave decelerates faster and the edge of the jet becomes visible at an earlier time (the jet break).  
The highest density model is again impacted by opacity at $150$~MHz, limiting its peak flux density to $0.03$~mJy and creating a plateau lasting until about 1 day.

All on-axis and isotropic cases produce a low flux, reaching at most 0.23~mJy at $150$~MHz ($0.34$~mJy at $863.5$~MHz) for the spherical high-density model at about 50 days (7 days) after the LIGO event.  
This would be a very difficult detection for any low-frequency radio array due to source confusion \citep{tingay13}.

\subsection{Off-axis normal short GRB}

Next, we model the afterglow as arising form an off-axis short GRB with a range of typical short GRB parameters.  
In all cases, we assume a jet with a half-opening angle of $\theta_j=10$~degrees, consistent with short GRB observations \citep{fong15}.
LIGO loosely constrains the inclination of the orbital plane of the binary black holes in 
GW150914, with the most likely value being an inclination of $\sim150$~degrees \citep{ligo16a}.
Assuming the jet axis is aligned with the binary orbital plane, this would place the jet axis $\sim30$~degrees away from our line of sight ($\theta_{obs}=30$~degrees).
We model observer angles of $20$, $30$, and $40$ degrees to cover a range of possible values.
We also model a range of external densities ($n_H=10^{-2}$, $10^{-1}$ and $10^{0}$~cm$^{-3}$) and isotropic energies within the jet to $E_{iso}=10^{51}$ or $10^{52}$~erg.
All off-axis models have an isotropic component with an energy of $1.8\times10^{49}$~erg to account for the {\it Fermi} GBM gamma-rays.  
In reality, the Earth-directed component need not be isotropic but could be, for example, the wing of a jet that declines rapidly outside of the jet core \citep{aloy05}.
For simplicity, we consider an isotropic component, responsible for early-time emission, and an off-axis top-hat jet, which creates a secondary peak and dominates late-time emission.

\begin{figure}
\centering
\includegraphics[width=0.7\columnwidth]{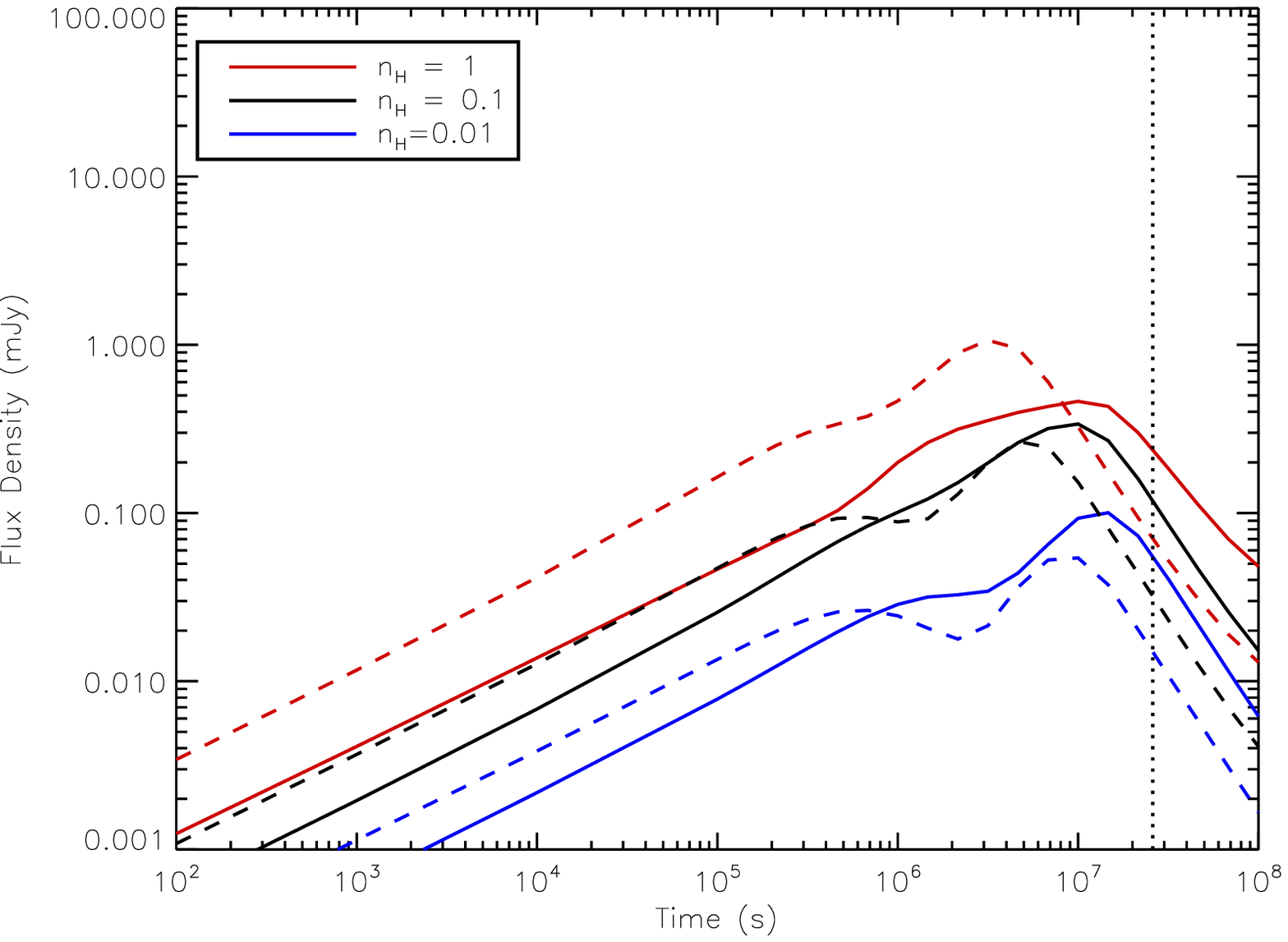}
\includegraphics[width=0.7\columnwidth]{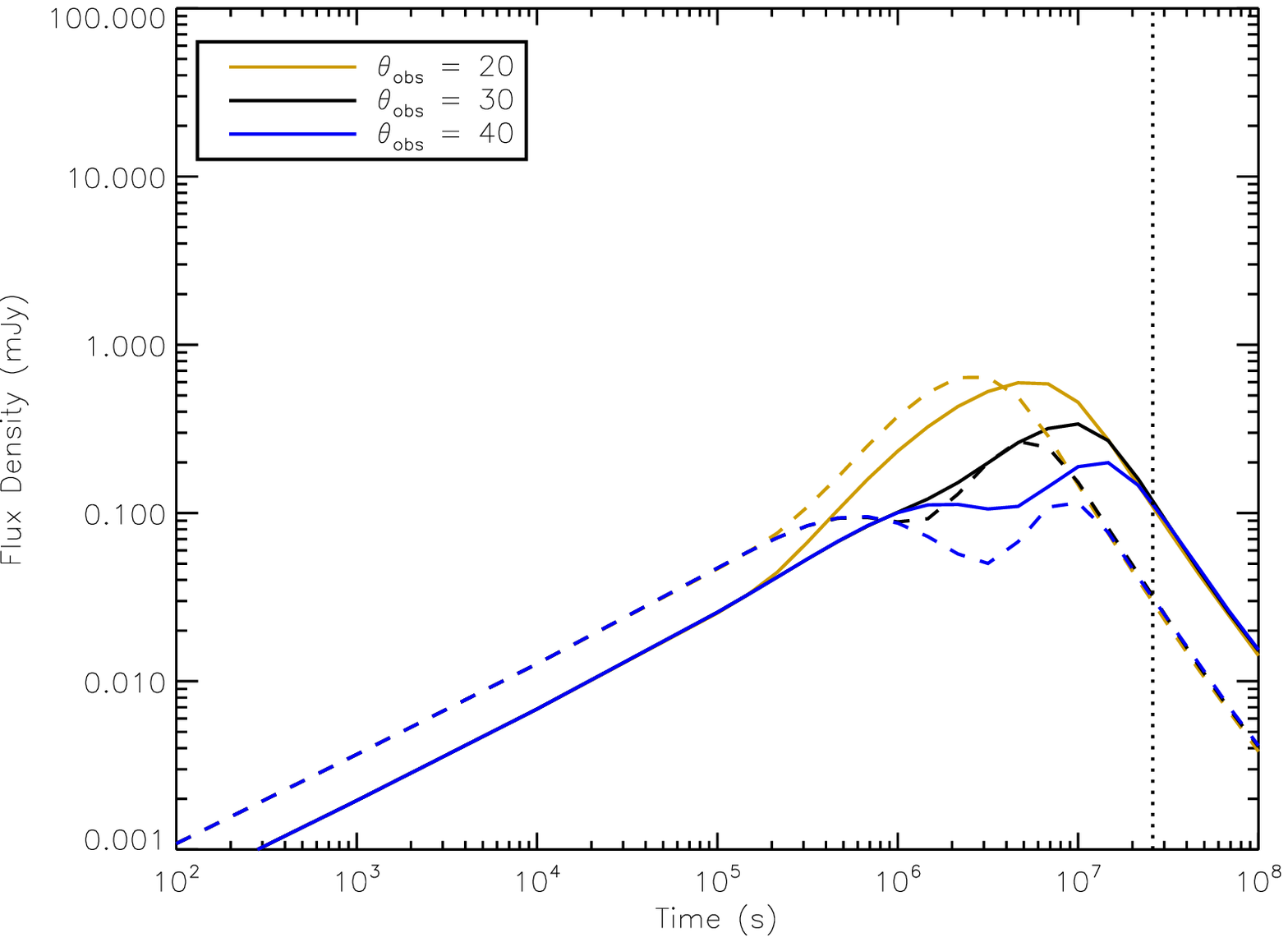}
\caption{
{\it Top}: Same as fig.~\ref{fig:sphere_jet}, but assuming an off-axis jet with $E_{iso}=10^{51}$~erg in addition to the isotropic component, with the observer at $\theta_{obs}=30$~degrees.
{\it Bottom}: Same as top panel, but assuming an external density of $n_H=0.1$~cm$^{-3}$ for all models, and an observer angle of $\theta_{obs}=20$ (orange), $30$ (black), and $40$ (blue) degrees.}
\label{fig:offaxis_n_theta}
\end{figure}

Fig.~\ref{fig:offaxis_n_theta} (top panel) shows our models for an $E_{iso}=10^{51}$~erg jet $30$~degrees off-axis, with 3 different external densities at $150$~MHz (solid lines) and $863.5$~GHz (dashed lines).  
The initial evolution follows the isotropic model cases (see fig.~\ref{fig:sphere_jet}).
However, after 3 days to 3 weeks, the jet decelerates sufficiently that it begins to become visible to an off-axis observer.
High external densities decelerate the jet faster, producing an earlier and brighter peak flux.
For an external density of $n_H=1$~cm$^{-3}$, the jet is still self-absorbed at $150$~MHz when it comes into view, reducing the peak flux and delaying the observed peak until 5 months vs. 1 month at $863.5$~MHz.
The self-absorbed case will also have a hard spectrum (slope of $\sim1.6$), while the other case  have a soft spectrum (slope of $\sim0$) at $150$~MHz.
The $n_H=0.1$ and $1$~cm$^{-3}$ models reach a peak flux densities of about $0.34$~mJy and $0.46$~mJy 4 to 5 months after the GRB.
All three cases would be easily detectable by ATCA at higher frequencies with a good localization.

Fig.~\ref{fig:offaxis_n_theta} (bottom panel) shows the results of varying the observer angle for models with $E_{iso}=10^{51}$~erg and an $n_H=0.1$~cm$^{-3}$.
Moving closer to the jet axis increase the observed peak flux and moves the peak earlier in time, from about 1 months to 2 months to 3 months at $863.5$~MHz for $\theta_{obs}=20$, $30$, and $40$~degrees, respectively.
At late times, once the entire jet is in view, the declining flux is nearly identical regardless of observer angle.

\begin{figure}
\centering
\includegraphics[width=0.7\columnwidth]{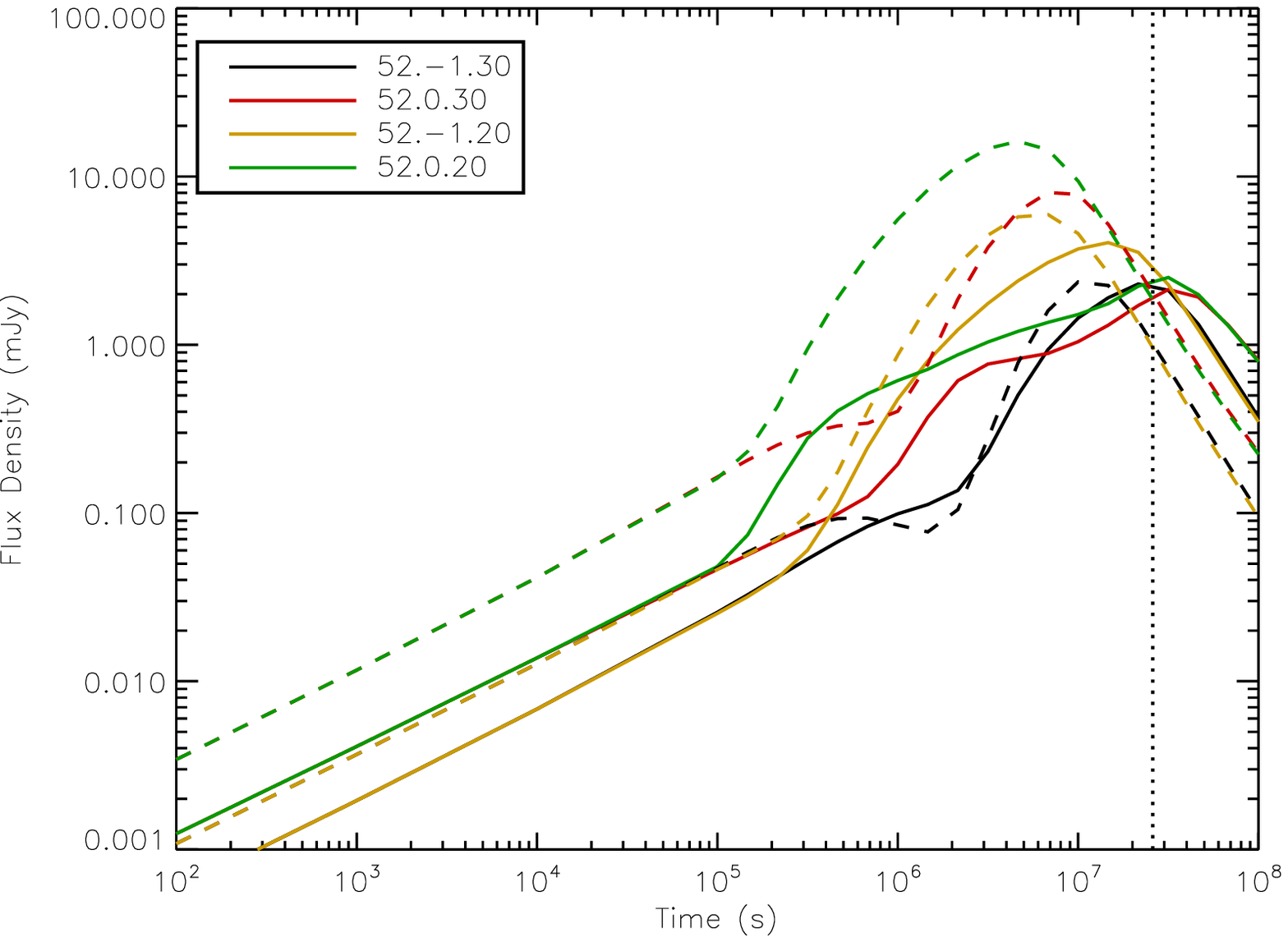}
\includegraphics[width=0.7\columnwidth]{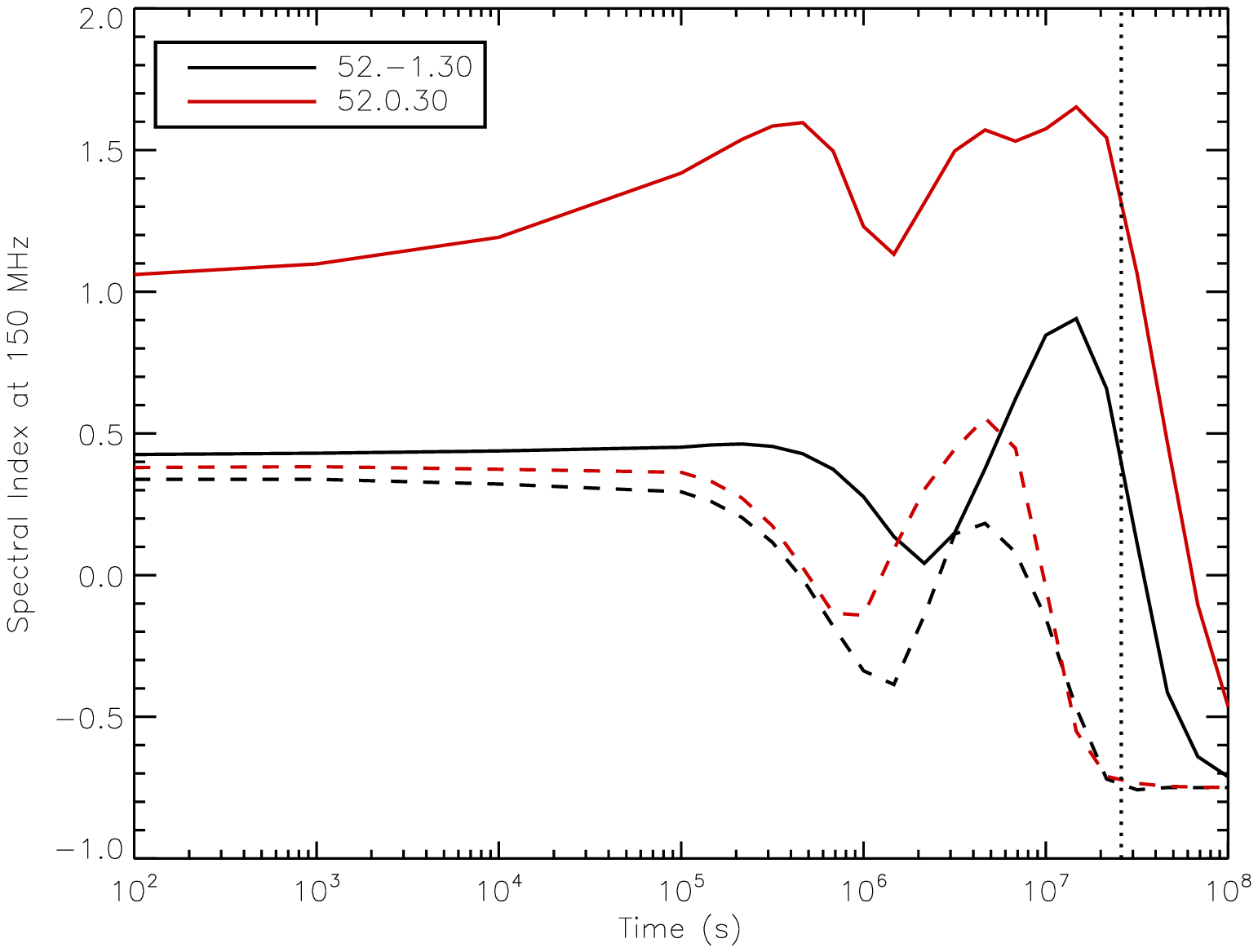}
\caption{
{\it Top}: Same as fig.~\ref{fig:offaxis_n_theta}, but showing afterglow models for $E_{iso}=10^{52}$~erg for $\theta_{obs}=30$~degrees with external densities of $n_H=0.1$~cm$^{-3}$ (black) and $1$~cm$^{-3}$ (red), and $\theta_{obs}=20$~degrees with external densities of $n_H=0.1$~cm$^{-3}$ (orange) and $1$~cm$^{-3}$ (green).  Names in legends correspond to table~\ref{table:models}.
{\it Bottom}: The spectral slope of radio emission averaged between $80$ and $300$~MHz (solid lines) and between $713.5$ and $1013.5$~MHz (dashed lines) for models with $E_{iso}=10^{52}$~erg, $\theta_{obs}=30$~degrees and external densities of $n_H =0.1$~cm$^{-3}$ (black) and $1$~cm$^{-3}$ (red).  
Dotted vertical lines are set 10 months after the initial event, corresponding to July 10, 2016 for GW150914, the approximate publication date of this paper.}
\label{fig:offaxis_E52}
\end{figure}

If the jet energy is $E_{iso}=10^{52}$~erg, fig.~\ref{fig:offaxis_E52} (top panel), the peak time moves later by a factor of two and the peak flux increases a factor of about 7 relative to models with $E_{iso}=10^{51}$~erg.
With a peak flux density of $2-4$~mJy at $150$~MHz 6 to 12 months after GW150914, even a GRB with these parameters would be at most marginally detectable with the MWA, though a source may continue to brighten for several months after the vertical dotted line.
The $863.5$~MHz flux density is up to $16$ mJy and should be detectable by ASKAP.  It would also be easily detectable by the ATCA given an afterglow candidate or sufficiently small search region.

At low frequencies the flux peak occurs near the time when the jet is becoming optically thin.  This leads to a large change in spectral slope near the peak, as seen in fig.~\ref{fig:offaxis_E52} (bottom panel).  The spectra just before the peak flux is reached becomes as hard as $F_{\nu}\propto\nu^{\sim1.6}$ in the high density case ($F_{\nu}\propto\nu^{\sim1}$ at medium density) before deceasing and ultimately falling to $F_{\nu}\propto\nu^{-0.75}$ at late times.  
At higher frequencies the jet becomes optically thin before the peak is reached, creating a less dramatic change in slope, from about $F_{\nu}\propto\nu^{1/3}$ before the peak to $F_{\nu}\propto\nu^{-0.75}$ at late times.
The change in spectral slope from hard to soft, particularly at low frequencies, could be used to find promising afterglow candidates for GW150914.

\subsection{X-ray and Optical Afterglow}

\begin{figure}
\centering
\includegraphics[width=0.7\columnwidth]{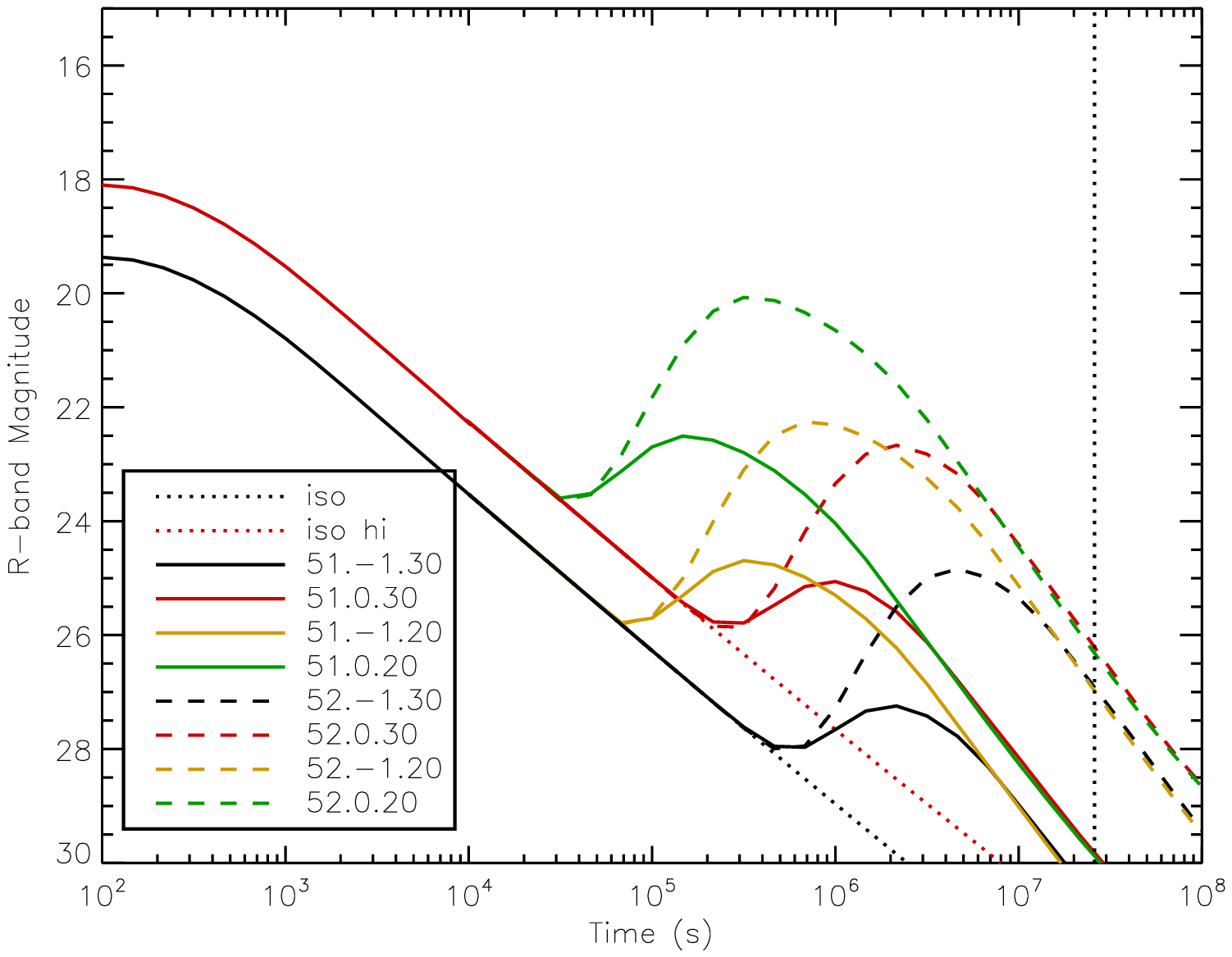}
\includegraphics[width=0.7\columnwidth]{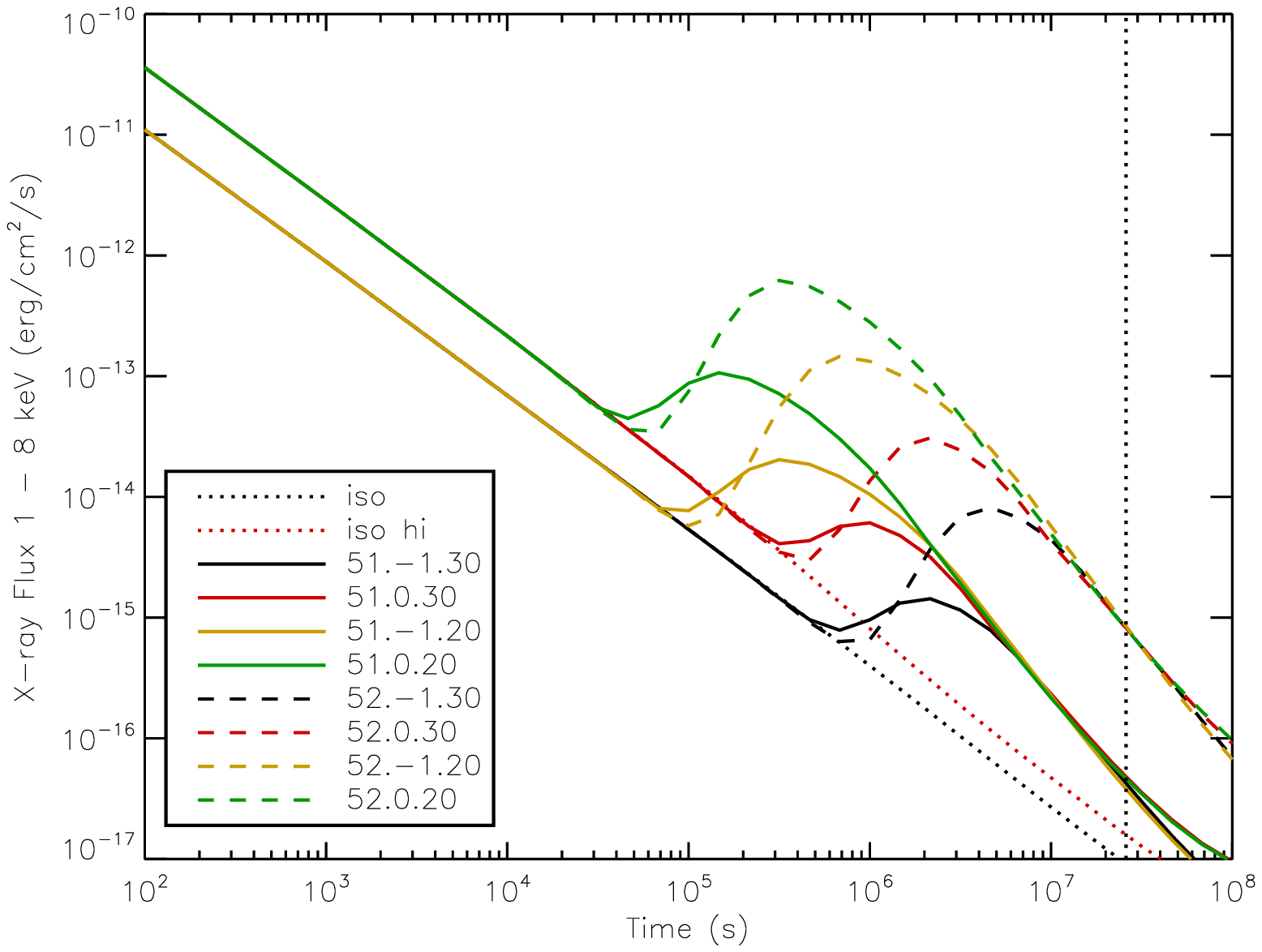}
\caption{
{\it Top}: Modeled afterglow R-band magnitude for selected models.  
{\it Bottom}: Modeled afterglow X-ray flux integrated from $1$ - $8$~keV for selected models.  Names in legends correspond to table~\ref{table:models}.}
\label{fig:optical_xray}
\end{figure}

The modeled R-band optical afterglow of select models are shown in fig.~\ref{fig:optical_xray} (top panel).
At early times, before the relativistic outflow has decelerated significantly, the after glow is dominated by the Earth-directed energy.  
Optical emission initially peaks at $100$s at about magnitude 18.1 (19.4) for an external density of $1$ ($0.1$)~cm$^{-3}$ and then rapidly fades.

For the off-axis models, there is a second peak when the jet comes into view.
This peak is only brighter than $22$nd magnitude for the most optimistic model parameters, e.g. a combination of high external density, high $E_{iso}$, and small observer angle.
However, under such favorable circumstances the afterglow of GW150914-GRB would have been detectable by Pan-STARRS at several days to months after the LIGO detection, if it was in Pan-STARRS field of view.
The late time optical luminosity is set by the energy of the jet, and by mid-July 2016 even the most optimistic models would be $26$th magnitude.
Our optical modeling does not include potential contributions from a non-relativistic supernova or kilonova counterpart.

In the X-ray (fig.~\ref{fig:optical_xray}, bottom panel), the afterglow is as bright as $3.6\times10^{-11}$~erg~cm$^{-2}$~s$^{-1}$ between $1$ and $8$~keV and is still brighter than $10^{-14}$~erg~cm$^{-2}$~s$^{-1}$ at 1 day in all models.
For the off-axis models, the X-rays re-brighten to above $2.6\times10^{-16}$~erg~cm$^{-2}$~s$^{-1}$, and to as much as $6.3\times10^{-13}$~erg~cm$^{-2}$~s$^{-1}$ in the brightest case.
This is below the detection limit of {\it Swift} LMC observations over a portion of the error box of GW150914 \citep{evans16}, and just at the detection limit of XMM slew observations \citep{troja16}.

At late times, the X-ray flux is mostly set by the $E_{iso}$ of the jet.
At the time this paper was published, the X-ray afterglow would be about $4\times10^{-16}$~erg~cm$^{-2}$~s$^{-1}$ for $E_{iso}=10^{51}$~erg and $10^{-15}$~erg~cm$^{-2}$~s$^{-1}$ for $E_{iso}=10^{52}$~erg.
If GW150914 were localized, an afterglow this bright would still be detectable by {\it Chandra} or XMM observations.

\section{Conclusions}
\label{sect:conclusions}

If GW150914-GBM is associated with the LIGO detection of GW150914, and assuming it is an off-axis short GRB with a relatively bright jet ($E_{iso}=10^{52}$~erg), its afterglow would be detectable {\bf NOW} (mid-July 2016) at $863.5$~MHz with ASKAP, but probably not at $150$~MHz with the MWA (see fig~\ref{fig:offaxis_E52}).  
Under these conditions, if an afterglow candidate were found, it would be easily observable at higher radio frequencies by the ATCA ($\sim1$~mJy at 1.4GHz), in X-rays by {\it Chandra} or XMM ($\sim10^{-15}$~erg~cm$^{-2}$~s$^{-1}$), and possibly with deep optical followup ($\sim26$th mag. in R-band).

If GW150914-GBM has a more typical short GRB energy of $E_{iso}=10^{51}$~erg, it would likely not be detectable by the MWA, reaching a peak flux density of at most $0.6$~mJy at $150$~MHz.
It may still be detectable with ASKAP at $863.5$~MHz, where the peak flux density is as high as $2$~mJy.
However, if an afterglow candidate or sufficiently small error box were identified such an afterglow would be observable by the ATCA ($>0.02$~mJy at $1.4$~GHz for all models) and possibly deep X-ray observations ($\sim4\times10^{-17}$~erg~cm$^{-2}$~s$^{-1}$) in mid-July 2016.

If instead GW150914-GBM is a very low luminosity short GRB, either isotropic or beamed, with $E_{iso}=1.8\times10^{49}$~erg, it was likely never bright enough to be observable without an immediate localization.  Only the higher frequency radio flux would be potentially detectable after the first few hours, at $\sim0.3$~mJy for a dense external medium.

Given the high rates of black hole mergers inferred from the LIGO detection of GW150914 \citep{abbott16b}, if such events do produce short GRBs, future events with slightly more favorable properties should be observable.  
Having a large GRB jet energy, high external density, jet axis closer to our line of sight, closer distance to Earth, or better initial localization would all favor the detection of an associated relativistic afterglow with follow-up radio, optical, and X-ray searches.

\section*{Acknowledgments}

The authors thank the anonymous referees for their insightful comments.
BJM was supported by an NSF AAPF under grant AST1102796, by the Aspen Center for Physics under NSF grant PHY-1066293, and by the NSF under grant AST1333514.
Dominic Ryan was supported as an REU student under NSF grant AST1004881.


\begin{thebibliography}{}

\bibitem[Abbott et al.(2016a)]{abbott16a} Abbott, B. P., Abbott, R., Abbott, T. D., et al. 2016, Phys. Rev. Lett., 116, 061102

\bibitem[Abbott et al.(2016b)]{abbott16b} Abbott, B. P., Abbott, R., Abbott, T. D., et al. 2016, ArXiv e-prints, arXiv:1602.03842

\bibitem[Aloy et al.(2005)]{aloy05} Aloy, M. A., Janka, H.-T., \& {M{\"u}ller}, E. 2005, \aap, 436, 273

\bibitem[GCN 18363()]{gcn18363} Bannister, K., et al. 2015, GCN Circ. 18363, http://gcn.gsfc.nasa.gov/gcn3/18363.gcn3

\bibitem[Blandford \& McKee(1976)]{blandford76} Blandford, R. D., \& McKee, C. F. 1976, Physics of Fluids, 19, 1130

\bibitem[Cannizzo et al.(2004)]{cannizzo04} Cannizzo, J. K., Gehrels, N., \& Vishniac, E. T. 2004, \apj, 601, 380

\bibitem[Connaughton et al.(2016)]{connaughton16} Connaughton, V., Burns, E., Goldstein, A., et al. 2016, ArXiv e-prints, arXiv:1602.03920

\bibitem[DeColle et al.(2012a)]{decolle12} De Colle, F., Granot, J., L\'opez-C\'amara, D., \& Ramirez-Ruiz, E. 2012, \apj, 746, 122

\bibitem[De Colle et al.(2012b)]{decolle12b} De Colle, F., Ramirez-Ruiz, E., Granot, J., \& L\'opez-C\'amara, D. 2012, \apj, 751, 57

\bibitem[Evans et al.(2016)]{evans16} Evans, P. A., Kennea, J. A., Barthelmy, S. D., et al. 2016, ArXiv e-prints, arXiv:1602.03868

\bibitem[Fong et al.(2015)]{fong15} Fong, W., Berger, E., Margutti, R., \& Zauderer, B. A. 2015, \apj, 815, 102

\bibitem[Granot et al.(2002)]{granot02} Granot, J., Panaitescu, A., Kumar, P., \& Woosley, S. E. 2002, \apjl, 570, L61

\bibitem[Granot et al.(1999)]{granot99} Granot, J., Piran, T., \& Sari, R. 1999, \apj, 527, 236

\bibitem[Hobbs et al.(2016)]{hobbs16} Hobbs, G., Heywood, I., Bell, M. E., et al. 2016, \mnras, 456, 3948

\bibitem[Leventis et al.(2012)]{leventis12} Leventis, K., van Eerten, H. J., Meliani, Z., \& Wijers, R. A. M. J. 2012, \mnras, 427, 1329

\bibitem[Li et al.(2016)]{li16} Li, X., Zhang, F.-W., Yuan, Q., et al. 2016, ArXiv e-prints, arXiv:1602.04460

\bibitem[Morsony et al.(2007)]{morsony07} Morsony, B. J., Lazzati, D., \& Begelman, M. C. 2007, \apj, 665, 569

\bibitem[Morsony et al.(2009)]{morsony09} Morsony, B. J., Workman, J. C., Lazzati, D., \& Medvedev, M. V. 2009, \mnras, 392, 1397

\bibitem[Panaitescu \& Kumar(2000)]{panaitescu00} Panaitescu, A. \& Kumar, P. 2000, \apj, 543, 66

\bibitem[Piran(2004)]{piran04} Piran, T. 2004, Reviews of Modern Physics, 76, 1143

\bibitem[Rossi et al.(2004)]{rossi04} Rossi, E. M., Lazzati, D., Salmonson, J. D., \& Ghisellini, G.
2004, \mnras, 354, 86

\bibitem[Savchenko et al.(2016)]{savchenko16} Savchenko, V., Ferrigno, C., Mereghetti, S., et al. 2016, ArXiv e-prints, arXiv:1602.04180

\bibitem[Smartt et al.(2016)]{smartt16} Smartt, S. J., Chambers, K. C., Smith, K. W., et al. 2016, ArXiv e-prints, arXiv:1602.04156

\bibitem[Soares-Santos et al.(2016)]{soares-santos16} Soares-Santos, M., Kessler, R., Berger, E., et al. 2016, ArXiv e-prints, arXiv:1602.04198

\bibitem[LIGO(2016)]{ligo16a} The LIGO Scientific Collaboration, \& the Virgo Collaboration. 2016, ArXiv e-prints, arXiv:1602.03840

\bibitem[Tingay et al.(2013)]{tingay13} Tingay, S. J., Goeke, R., Bowman, J. D., et al. 2013, PASA, 30,
e007

\bibitem[Troja et al.(2016)]{troja16} Troja, E., Read, A. M. Tiengo, A. \& Salaterra, R. 2016, \apjl, 822, L8

\bibitem[van Eerten et al.(2010)]{vanEerten10} van Eerten, H., Zhang, W., \& MacFadyen, A. 2010, \apj, 722, 235

\bibitem[van Eerten \& MacFadyen(2011)]{vanEerten11} van Eerten, H. J., \& MacFadyen, A. I. 2011, \apjl, 733, L37

\bibitem[van Eerten \& Wijers(2009)]{vanEerten09} van Eerten, H. J., \& Wijers, R. A. M. J. 2009, \mnras, 394, 2164

\bibitem[Wanderman \& Piran(2015)]{wanderman15} Wanderman, D., \& Piran, T. 2015, \mnras, 448, 3026

\bibitem[Waxman(2004)]{waxman04} Waxman, E. 2004, \apj, 602, 886

\end{thebibliography}


\end{document}